\documentclass[conference]{IEEEtran}

\IEEEoverridecommandlockouts
\ifCLASSINFOpdf
\usepackage{cite}
\usepackage{amsmath,amssymb,amsfonts}
\usepackage{algorithmic}
\usepackage{graphicx}
\usepackage{textcomp}
\usepackage{xcolor}
\usepackage{caption}
\usepackage{comment}
\usepackage{enumitem}
\usepackage{graphicx}
\else
\fi
\def\BibTeX{{\rm B\kern-.05em{\sc i\kern-.025em b}\kern-.08em
    T\kern-.1667em\lower.7ex\hbox{E}\kern-.125emX}}
\begin{document}

\title{Evolution of Ethereum: A Temporal Graph Perspective \\
}
\author{\IEEEauthorblockN{Qianlan Bai$^{\dag}$, Chao Zhang$^+$, Yuedong Xu$^+$, Xiaowei Chen$^*$, Xin Wang$^{\dag}$}
	\IEEEauthorblockA{$^+$ School of Information Science and Engineering, Fudan University}
	\IEEEauthorblockA{$^{\dag}$ School of Computer Science and Technology, Fudan University}$^{*}$School of Computer Science and Engineering,
		The Chinese University of Hong Kong
	}
\maketitle

\begin{abstract}
	Ethereum is one of the most popular blockchain systems that supports more than half a million transactions every day and fosters miscellaneous decentralized applications with its Turing-complete smart contract machine. 
	Whereas it remains mysterious what the transaction pattern of Ethereum is and how it evolves over time. In this paper, we study the evolutionary behavior of Ethereum transactions from a temporal graph point of view. 
	We first develop a data analytics platform to collect external transactions associated with users as well as internal transactions initiated by smart contracts. Three types of temporal graphs, user-to-user, contract-to-contract and user-contract graphs, are constructed according to trading relationship and are segmented with an appropriate time window. 
	We observe a strong correlation between the size of user-to-user 
	transaction graph and the average Ether price in a time window, while no evidence of such linkage is shown at the average degree, average edge weights and average triplet closure duration. The macroscopic and microscopic burstiness of Ethereum transactions is validated. We analyze the Gini indexes of the transaction graphs and the user wealth in which Ethereum is found to be very unfair since the very beginning, in a sense, ``the rich is already very rich''.

\end{abstract}

\section{Introduction}

Blockchain is an emerging technology that has the potential to revolutionize many traditional industries. 
As the first generation blockchain system, Bitcoin has proved that the global consensus can be achieved 
with no central authority or trusted third party. Since then, numerous efforts have been devoted to the 
design of more efficient and multifunctional blockchain systems. Ethereum is the second largest blockchain 
platform whose market value has reached 1573 millions dollors on Jan 1, 2020. Around half a million transactions are conducted on the Ethereum platform
everyday. 
More importantly, it supports smart contract that is a collection of codes and data residing at a specific address on Ethereum, and running on Ethereum virtual machine (EVM). The smart contracts, together with the improved transaction speed, propels the wide adoption of Ethereum and foster versatile decentralized applications.

The boom of Ethereum has aroused great interests in the understanding of its social interactions. 
The activities of its users are encapsulated in the blocks where each transaction inside a block 
contains the sending 
and receiving addresses, the value transferred and so on. As an open distributed ledger, Ethereum allows any user to store all the transactions. Therefore, analyzing transaction data provides a crucial way to know the fundamental properties of 
Ethereum and its development. A commonly used approach is to construct transaction graphs, in which 
each address is expressed as a node, and a trade between two addresses is expressed as a (directed) edge.

Existing studies on Ethereum transaction graphs can be roughly categorized into two types. One is the analysis of 
static graphs. Chen \emph{et.al} \cite{ref:infocom} characterized the activity patterns of money transfer, contract creation and contracts invocation using different graphs. Kiffer \emph{et.al} in \cite{ref:Analyzing Ethereum's Contract Topology} 
investigated how Ethereum smart contracts are created and how users and smart contracts interact with one another. 
A quantitative study on the transfer graph of Ethereum tokens was conducted in \cite{ref:Measuring ethereum-based erc20 token networks}. 
The other is the data-driven problem solving in which the abnormal event detection is the focus of interest. 
Chen \emph{et.al} \cite{ref:ponzi} designed a machine learning approach using 
account-level and bytecode-level features to classify Ponzi addresses. Chen \emph{et. al} \cite{ref:chen} studied the 
the inconsistent token behaviors with regard to ERC-20 of Ethereum, and design a software toolkit to 
inspect transanctions sent to the deployed tokens. 
In the literature, Ethereum and its transaction graphs are usually analyzed as a static network, while its evolutionary 
behavior is largely overlooked.

In this paper, we study a few fundamental questions regarding Ethereum: \emph{How does Ethereum transaction pattern 
	evolve over time, and how is it affected by Ether price?} 
To this goal, we first develop a data analytics platform that parses archive node blocks to extract Ethereum external transactions, modifies client program to record internal transactions initiated by smart contracts, and crawls Etherscan 
as well as web forums. The transactions collected starts from July 30th (the birth date of Ethereum) to Feb 9th, 2019. We then construct three graphs, namely user-to-user graph (UUG), contract-to-contract graph (CCG) 
and user-contract graph (UCG). UUG characterizes the trading relationship among externally 
owned accounts (EoA), CCG captures 
the complex function invocations among smart contracts, and UCG is a bipartite graph capturing the transactions between 
EoAs and smart contracts. In order to understand the evolutionary behaviors of Ethereum transactions, we chop each 
transaction graph into a sequence of temporal graphs with a carefully chosen time window. Both the sliding 
and the incremental time window are considered. 

Our graph analysis on UUG is carried out from three perspectives. Firstly, we measure the degree distribution, edge weights and local graph structure (i.e. 3-node motif) that evolve over time. Secondly, the macroscopic burstiness that captures the aggregation extent of transactions made by nodes, and the microscopic burstiness that characterizes the 
inter-transaction time distribution is also quantified. Third, the Gini indexes of degree, transaction and balance (wealth) distributions are computed for the purpose of verifying ``the rich gets richer'' effect. In addition, 
we analyze the internal transaction behaviors with a temporal graph analysis on CCG and analyze the interaction between 
EOAs and smart contracts using UCG. 

Our major observations are briefly summarized as below:

\begin{itemize}[fullwidth,itemindent=0em]
	
	\item 1. The size of UUG, the total number and the total value of transactions in UUG experience three stages that are 
	consistent with the dynamics of Ether price. The average number and the average value of transactions  
	in each node and each edge decreases over time, which shows little evidence that this trend is Ether price-related. 
	
	\item 2. The global clustering coefficient of UUG is very small and it decreases over time. 
	The correlation between Ether price and the local graph structure such as the proportion of closed triplets and the 
	average closure time is not observed.

	\item 3. For most of nodes, their transactions are concentrated on a short duration of their active periods. 
	The inter-transaction time intervals varies considerably, exhibiting a certain degree of burstiness.

	\item 4. The distribution of degree, transaction and wealth of nodes are always unfair since the genesis of Ethereum.

	
	\item 5. The development of smart contracts can be also divided into three stages. In most of time windows, the number of 
	smart contracts created by EOAs is of the same magnitude as that created by other contracts, and the smart contracts are 
	usually invoked by EoAs. 
\end{itemize}
\section{Datasets and Basic Definitions}
\label{sec:dataset}

In this section, we describe our data collection process and data analytic platform. The construction of transaction graphs 
is introduced in detail.

\subsection{Overview of Datasets}

We collect all the Ethereum transactions spanning from July 30th, 2015, the birth date of Ethereum, to February 9th, 2019. 
The total number transactions is around \textbf{\emph{410 millions}}, and the size of our dataset is around 100 GB. 
Each transaction records the following items: \emph{\{Block ID, Transaction Hash, Sender, Receiver, Transaction Value\}}. 
The sender and the receiver are actually addresses known as public keys. 
We also collect the balance of each address at all time. 
Additional information regarding Ethereum ecology is also crawled from public forum, 
including 47 Mining pool accounts, 146 Exchange accounts, 
31 donation accounts, 170 ICO-wallets, 2884 phishing accounts and 184 Ponzi accounts \cite{ref:ponzi}. 

\subsection{Data Collection Procedure}
As a prominent feature of public Blockchain, all the transactions are traceable. In Bitcoin, each transaction relevant to monetary 
balance
can be fetched by examining the blocks. Ethereum, empowered by the invention of smart contract, unfortunately does not 
store complete information of all transactions, thus making the data collection more difficult. We leverage the following methods 
to collect transactions. 

\noindent\textbf{Parsing Archive Node.} To acquire transaction data, one needs to launch an Ethereum node 
as a \emph{Geth} client. The Geth client provides four synchronization methods: light, fast, full and archive modes. The light mode 
only downloads the header of blocks; the fast and the full modes further download the body of blocks, with gentle 
difference in the validation process; the archive mode enables the client to retain all the history data. 
We choose the archive mode and which occupies 3TB disk space. 
Note that the raw data in an Ethereum node is stored in binary. A powerful library Web3 is used to convert the raw data into 
readable text format. The APIs, \textit{getBlock}, \textit{getTransaction} and \textit{gethBalance}, are utilized to extract transaction information. Since 
the external accounts and the smart contract accounts are all stored as ``meaningless'' Hash addresses, 
we resort to the API \textit{getCode} to acquire the operation code of each address. If the return of \textit{getCode} function is 
``0x'', the corresponding account is an external one, and vice versa.

\noindent\textbf{Modifying Ethereum Client.} The internal transactions initiated by smart contracts are not recorded on blockchain and therefore unavailable through parsing a standard Ethereum node. 
Smart contracts are specially designed programs running on Ethereum Virtual Machine (EVM), and their operations will 
invoke some functions. Hence, to acquire the internal transactions, we slightly modify the transaction-related functions of EVM \emph{Geth} client including ``call()'', ``callcode()'' and ``create()'', and establish an 
private chain. All the external transactions are executed chronologically on the chain so as to simulate 
what has happened in the genuine Ethereum system. 

\noindent\textbf{Crawling Etherscan and Web Forums.} 
The Ethereum node does not record the information regarding the Ether price and identity of some known accounts (e.g. exchanges, wallets and frauds). 
We collect these informations from Etherscan (a comprehensive Ethereum explorer) and open forums. 
\subsection{Construction of Transaction Graphs}
Our purpose is to uncover the fundamental properties of Ethereum transactions in 
a network perspective enables us to understand the compound interactions among users. 
In light of the diverse address types in Ethereum, we propose to construct three graphs: 
user-to-user graph (\textbf{UUG}), contract-to-contract graph (\textbf{CCG}) and user-contract graph (\textbf{UCG}). 

In UUG graph, each Externally Owned Account (EOA) is a node and two nodes form a \emph{directed} edge if a transfer 
of Ether between them happens. 
The UUG captures the transaction patterns among Ethereum users, that is, the most important aspect of Ethereum as 
a cryptocurrency. 
Similarly, the CCG graph is constructed in the same way as the UUG graph except that each node is a smart contract, and 
the Ether transfer is substituted by the function ``create", ``call" or transfer between pair-wise smart contracts. 
The CCG graph demonstrates the transaction relationship that is triggered by ``codes'' automatically instead of human beings 
manually. 
The UCG graph is a bipartite graph where one side consists of only EOAs and the other side consist of only smart contracts. 
It reveals how the smart contracts are created, used by EOAs. Also, it reveals how the money transfer from contracts to EOAs. 
The basic statistics of UUG, CCG and UCG are expressed in Table \ref{tab1e_defination}. 
\begin{table}[h]
	\vspace{-0.3cm}
	\caption{Sizes of Transaction Graphs.}
	\centering 
	\begin{tabular}{|c|c|c|c|c|c|c|c|c|c|}
		\hline  
		{}&{UUG}&{UCG}&{CCG}\\
		\hline  
		{type of nodes}&{user}&{user$\backslash$contract}&{contract}\\
		\hline  
		{\#of nodes}&$41722479$&$26605865$&$5971038$\\
		\hline  
		{\#of edges}&$89194399$&$50792582$&$6148867$\\
		\hline 
		{\#of transactions}&$192300454$&$207448032$&$13470886$\\
		\hline 
	\end{tabular}
	\label{tab1e_defination}
	\vspace{-0.3cm}
\end{table}	

The static transaction graphs are insufficient to reveal the evolution of Ethereum system. This motivates us to 
construct temporal graphs, in which the whole measurement period is subdivided into a number of time windows. 
Two types of time windows are considered, the sliding window and the incremental window. 
We observe that more than 70\% of nodes have a lifetime (the interval between the first and last transaction time in our dataset) below 180 days, and thus setting the sliding 
window to be 180 days is suitable to evaluate the graph dynamics. 
The sliding window is shifted with a granularity of 45 days (i.e. $1/4$ of this time window) that is useful to 
compare the graphs at different stages. The incremental window expands from 180 days to 1260 days with the same granularity. 
Although the time window is a hyper parameter, it is carefully selected and its choice does not 
influence the main conclusions drawn in this work. 
	\section{Temporal UUG Network Analysis }
\label{sec:uug_temporal}

In this section, we investigate the basic properties of UUG temporal network including 
node/edge distributions, burstiness of transactions and local interactions as time evolves.

\subsection{Elementary Graph Size}

\begin{figure}[h]
	\vspace{-0.3cm}
	\begin{minipage}[h]{0.24\textwidth}
		\centering
		\setlength\abovecaptionskip{-0.5pt}
		\setlength\belowcaptionskip{-1pt}
		\includegraphics[width=1\textwidth]{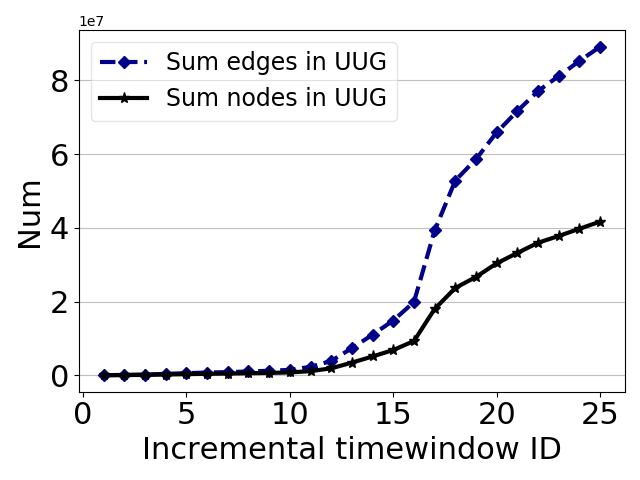}
		\caption{Size evolution in UUG with incremental window.}
		\label{fig:UUG size incremental}
	\end{minipage}
	\begin{minipage}[h]{0.24\textwidth}
		\centering
		\setlength\abovecaptionskip{-0.5pt}
		\setlength\belowcaptionskip{-1pt}
		\includegraphics[width=1\textwidth]{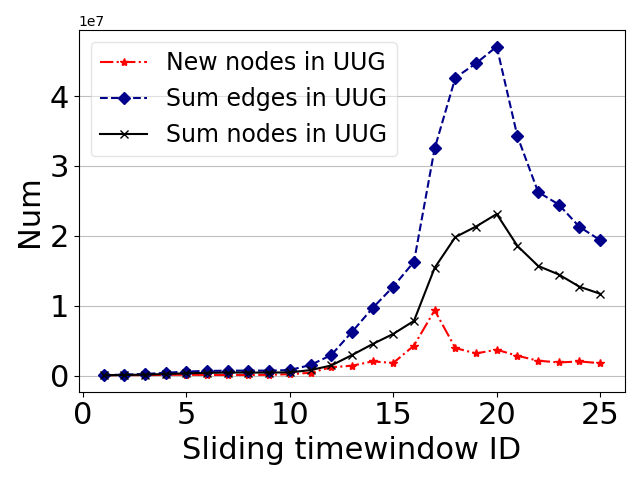}
		\caption{Size evolution in UUG with sliding window.}
		\label{fig:UUG size sliding}
	\end{minipage}
	\vspace{-0.3cm}
\end{figure}

Our measurement tour begins with analysis of UUG size. Figure \ref{fig:UUG size incremental} illustrates the number of nodes and edges 
in the incremental graphs. At the early stage of Ethereum, the quantities of the nodes and edges see a very gentle 
growth. Since March 17, Ethereum has experienced an extreme flash crowd of nodes and edges. 
In the last several time windows, the growth of UUG slows down, but still at a noticeable rate. 
We show the numbers of active nodes, active edges and newly created nodes in different sliding graphs in Figure \ref{fig:UUG size sliding}. 
The new nodes refer to those whose first transaction occurs within the current sliding window. 
From this figure, one can easily differentiate the development of Ethereum into three stages: ``slow start'' with slow increase
(March 15, 2015$\sim$March 15, 2017), ``outbreak'' with rapid increase (March 16, 2017$\sim$June 15, 2018), and ``fever abatement''  with considerable reduction (June 16, 2018$\sim$ ???). 
Comparing with Figure \ref{fig:Ether price}, we conjecture that the growth of UUG is heavily affected by the Ether price: the increase of Ether 
price attracting more 
transactions, and the drop of Ether price causing reduction in the transactions. 

\begin{figure}[h]
	\vspace{-0.3cm}
	\begin{minipage}[h]{0.24\textwidth}
		\centering
		\setlength\abovecaptionskip{-0.5pt}
		\setlength\belowcaptionskip{-1pt}
		\includegraphics[width=1\textwidth]{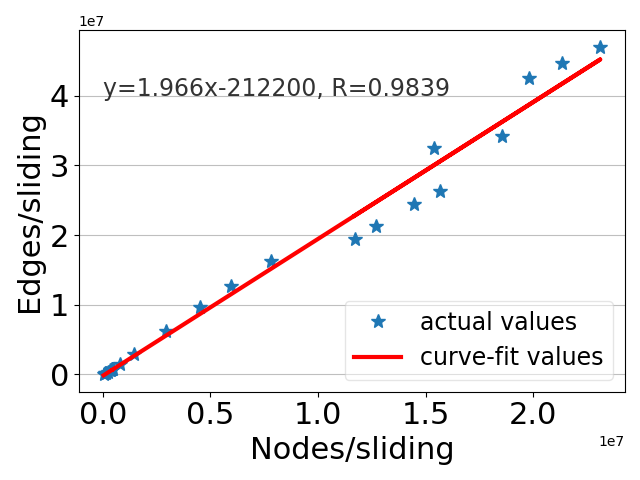}
		\caption{Relationship between nodes and edges in sliding graphs.}
		\label{fig:UUG node and edge fit sliding}
	\end{minipage}
	\begin{minipage}[h]{0.24\textwidth}
		\centering
		\setlength\abovecaptionskip{-0.5pt}
		\setlength\belowcaptionskip{-1pt}
		\includegraphics[width=1\textwidth]{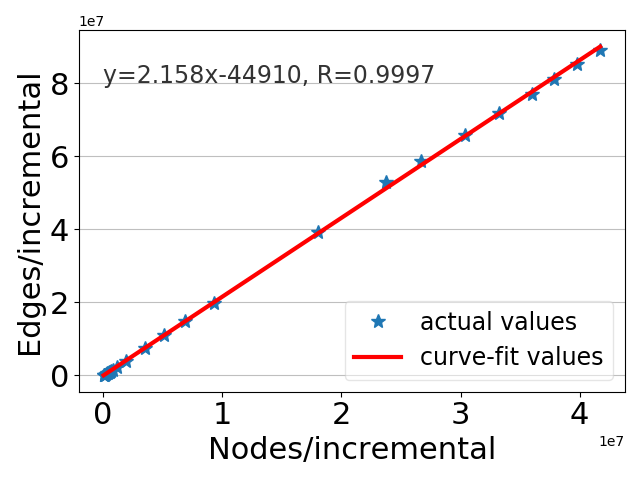}
		\caption{Relationship between nodes and edges in incremental graphs.}
		\label{fig:UUG node and edge fit incremental}
	\end{minipage}
	\vspace{-0.3cm}
\end{figure}

The correlation between the number of nodes and edges in different time windows is studied. In general,
as the random graphs evolve over time, they follow a version of the relation
\begin{equation}
\setlength\abovedisplayskip{1pt}
\setlength\belowdisplayskip{1pt}
\mathbf e(t)\propto n(t)^{\alpha}\label{eq:node and edge reference}
\end{equation}
where $e(t)$ and $n(t)$ denote the number of edges and nodes of graph at time $t$, 
and ${\alpha}$ is an exponent that lies strictly between 1 and 2 \cite{ref:node and edge}. We show such correlations of the 
sliding graphs window and the incremental window graphs in Figure \ref{fig:UUG node and edge fit sliding} and \ref{fig:UUG node and edge fit incremental} respectively. 
Note that the R-square value is very close to 1, indicating an excellent fit of our models. 

{ \textbf{Observation 1.}} The attractiveness of Ethereum to new users grows firstly and then it weakens, the trend is consistent with trend of Ether price. 		
There is a linear relationship between nodes and edges in the sliding window and the incremental window graphs, which 
complies with the basic property of random graph. 
	\subsection{Degree Distribution and Degree Properties}
We evaluate the degree distribution of UUG as an unweighted graph in order to look into the interactions among 
Ethereum addresses. Each edge is then associated with a weight that represents the frequency of transactions 
or the amount of Ethers flowing on this edge.

\subsubsection{Degree Distribution}
\begin{figure}[h]
	\vspace{-0.4cm}	
	\begin{minipage}[h]{0.24\textwidth}
		\centering
		\setlength\abovecaptionskip{-0.5pt}
		\setlength\belowcaptionskip{-1pt}
		\includegraphics[width=1\textwidth]{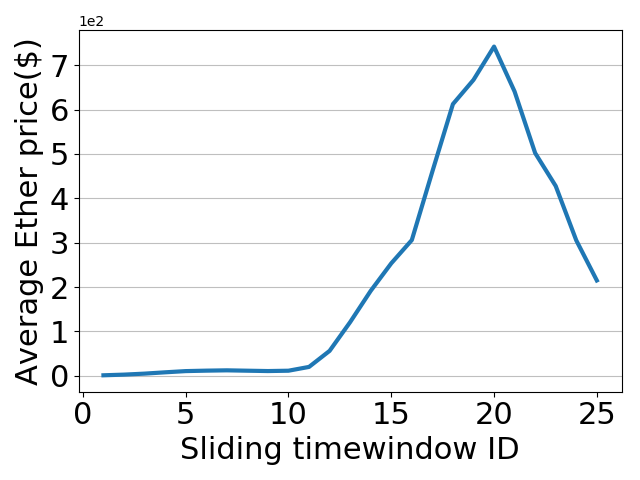}
		\caption{Average Ether price.}
		\label{fig:Ether price}
	\end{minipage}
	\begin{minipage}[h]{0.24\textwidth}		
		\centering
		\includegraphics[width=1\textwidth]{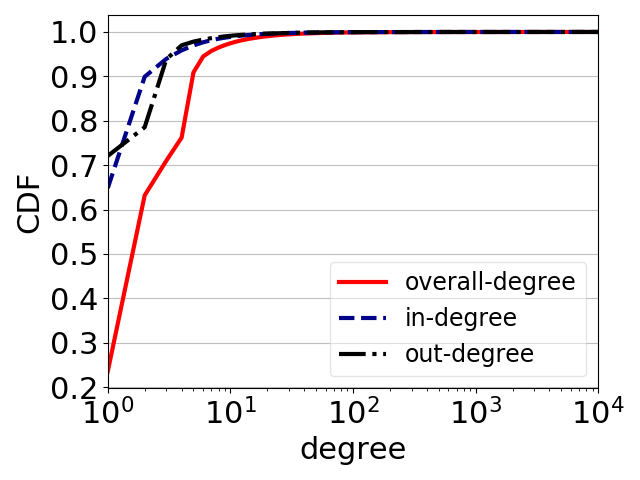}
		\caption{Degree distribution.}
		\label{fig:UUG degree distribution}
	\end{minipage}
	\vspace{-0.5cm}
\end{figure}

The degree distribution of a network is defined to be the fraction of nodes in the network that have a certain number of edges. 
We illustrate the in-degree, out-degree and overall-degree distributions of the up-to-date UUG in our measurement in Figure \ref{fig:UUG degree distribution}. Despite of the vast amount of transactions, nearly 23.58\% of nodes have transactions with only one address and 97.45\% of nodes have transactions with less than ten addresses. In contrast, a tiny fraction of nodes transfer Ethers to and from a large number of other nodes so that the degree distribution is highly skewed and exhibits a heavy-tail pattern. 
For instance, the most powerful node has a degree of 2605515, accounting for 6.24\% of all the nodes. 
We further use the crawled information to show the identities of top 20 nodes with the most edges 
in Table \ref{table:top degree}. The exchange nodes 
play a key role in the UUG, and they are similar to the hub nodes of on-line social networks\cite{ref:social network}. 

A simple curve fitting is used to approximate the degree distribution as the following
\begin{equation}
\setlength\abovedisplayskip{1pt}
\setlength\belowdisplayskip{1pt}
\mathbf{CDF_{degree}}=1-0.79*degree^{-1.23}\label{eq:UUG degree model}
\end{equation}
with the R-square to be 0.97. 

\begin{table}[h]
	\vspace{-0.2cm}
	\centering  
	\caption{Top 20 nodes with most degrees.}
	\begin{tabular}{|c|c|c|c|c|}
		\hline  
		{}&{Exchange}&{Mining pool}&{Ordinary node}&{ICO-wallet}\\
		\hline  
		{in-degree}&{$13$}&{$0$}&{$6$}&{$1$}\\
		\hline  
		{out-degree}&{$16$}&{$2$}&{$1$}&{$1$}\\
		\hline  
		{degree}&{$17$}&{$2$}&{$0$}&{$1$}\\
		\hline 
	\end{tabular}
	\label{table:top degree}
	\vspace{-0.4cm}
\end{table}

To explore the dynamics of UUG's degree distribution, we plot the average degree at the different sliding windows. 
One can see that the average degree becomes slightly higher between the $11^{th}$ and the $21^{st}$ sliding windows. 
Such an increase coincides with the period of the sudden increase of Ether price. Unfortunately, the prosperity of 
Ether market does not bring an obviously denser transaction graph. 

\begin{figure}[h]
	\vspace{-0.5cm}
	\begin{minipage}[h]{0.24\textwidth}
		\centering
		\includegraphics[width=1\textwidth]{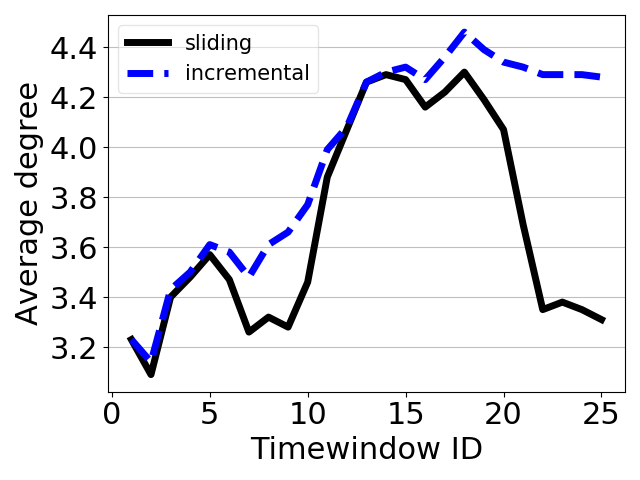}
		\caption{Average degree evolution in UUG.}
		\label{fig:UUG average degree}
	\end{minipage}
	\begin{minipage}[h]{0.24\textwidth}
		\centering
		\includegraphics[width=1\textwidth]{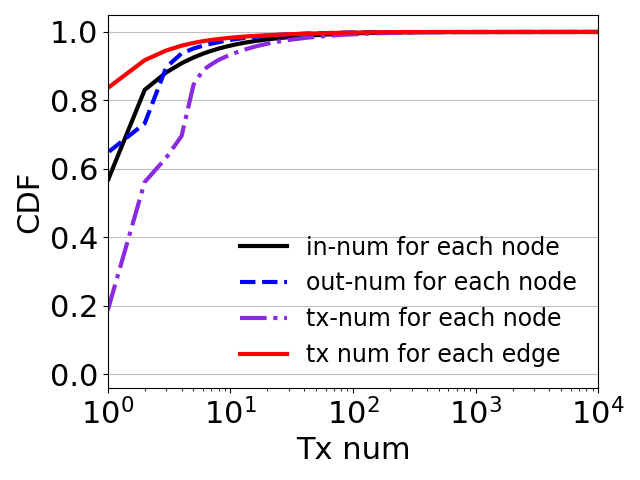}
		\caption{Transaction number distribution in UUG.}
		\label{fig:UUG num node edge}
	\end{minipage}
	\vspace{-0.5cm}
\end{figure}

\noindent\textbf{Observation 2:} Ethereum transaction graph is sparse and highly skewed. A tiny amount of nodes (e.g. exchange nodes or mining pools) have very large degrees, while the overwhelming majority of nodes only have transactions with several others. The increase of Ether price does not change the degree distribution remarkably. 

\subsection{Statistics on Edge Weights}
The degree distribution, though crucial to reveal the UUG structure, does not tell how frequently 
a transaction happens and how much money flows on an edge. 
Therefore, we measure the dynamics of the number of transactions and the accumulated money transfer of all
the edges in UUG. 

\begin{figure}[!h!t]
	\vspace{-0.3cm}
	\begin{minipage}[h]{0.24\textwidth}
		\centering
		\includegraphics[width=1\textwidth]{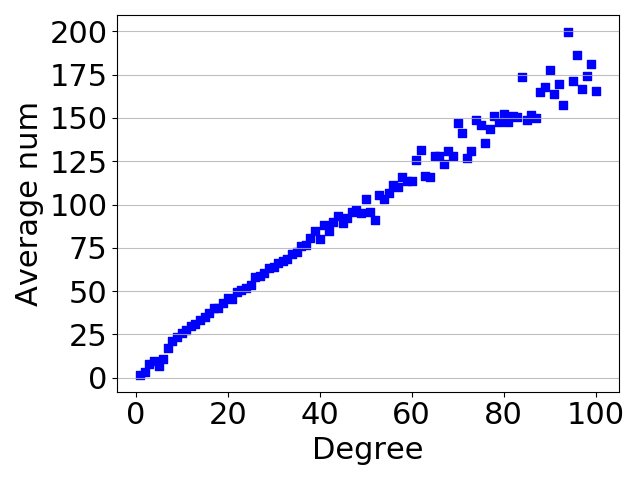}
		\caption{Average number of transactions versus degree.}
		\label{fig:UUG avg num degree}
	\end{minipage}
	\begin{minipage}[h]{0.24\textwidth}
		\centering
		\includegraphics[width=1\textwidth]{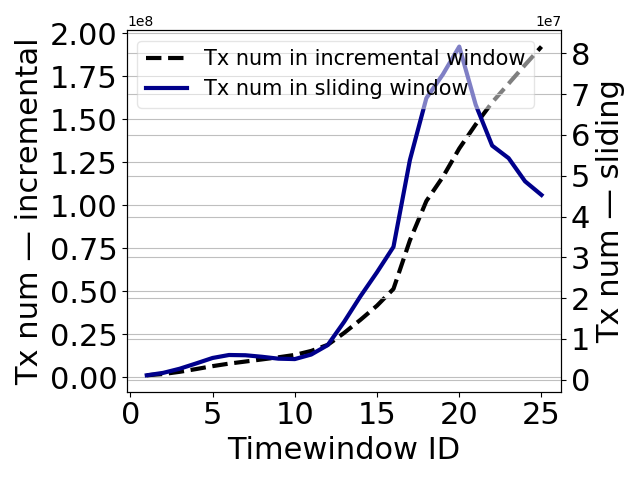}
		\caption{Number of transactions evolution in UUG.}
		\label{fig:UUG sumnum}
	\end{minipage}
	\vspace{-0.5cm}
\end{figure}

We first calculate the distribution of the number of transactions on each node and each edge separately in Figure \ref{fig:UUG num node edge}. 
Around 83.63\% of edges have only a one-time transaction, and around 18.69\% of nodes trade only once. 
Both of them leave heavy tails in their cumulative probability distributions. 
Note that the transactions of a node may come from the frequent Ether transfers on a few edges, or the infrequent 
transfers on many edges. We hereby explore the correlation between the edge distribution and the 
per-node transaction distribution. 
Figure \ref{fig:UUG avg num degree} plots the relationship between the degree and average number of transactions, in which an almost linear 
correlation is observed (around 99.9\% of nodes are included). 

We want to understand the trend that Ethereum transactions evolve. Figure \ref{fig:UUG sumnum} demonstrates the number of transactions with both the sliding and the incremental windows. 
Here, $x$-coordinate is the sequence number of the time window, and left $y$-axis is the total number of transactions 
in different incremental time windows, right $y$-coordinate is that in different sliding time windows. 
When examining the transaction number in the sliding windows, we observe the similar three-stage dynamics: 
``slow start'', ``outbreak'' and ``fever abatement''. After reaching the peak at the twentieth sliding window, 
the number of transactions experiences a drastic decrease, meaning that the Ethereum users tend to be less active now. 

\begin{figure}[!h!t]
	\vspace{-0.4cm}
	\begin{minipage}[!h]{0.24\textwidth}
		\centering
		\includegraphics[width=1\textwidth]{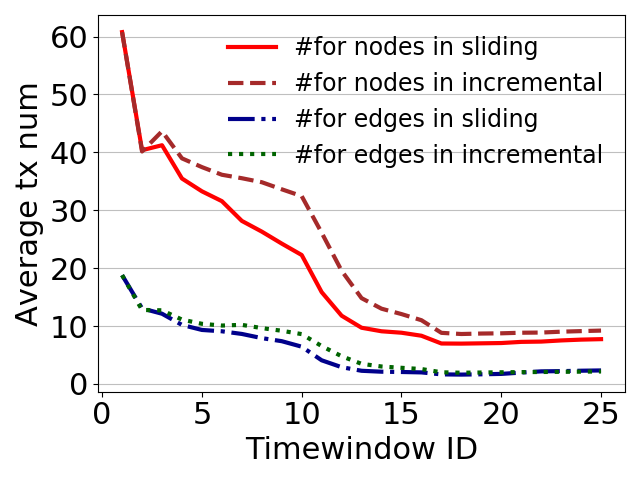}
		\caption{Average number of transaction in UUG with different time windows.}
		\label{fig:UUG average num}
	\end{minipage}
	\begin{minipage}[h]{0.24\textwidth}
		\centering
		\includegraphics[width=1\textwidth]{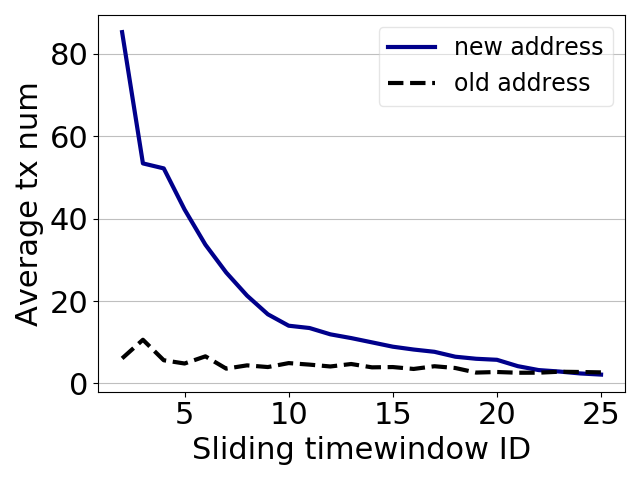}
		\caption{Average number of transactions of new and old address.}
		\label{fig:UUG avgnum new old}
	\end{minipage}
	\vspace{-0.5cm}
\end{figure}


Figure \ref{fig:UUG average num} shows the number of transactions on each node or each edge 
where both the sliding window and the incremental window are considered. 
The average number of transactions in all curves almost always decrease over time. 
A counterintuitive observation is that the average number of transactions becomes smaller and smaller. 
To uncover the reason, we differentiate the nodes into two groups, the old nodes and the newly created nodes, in Figure \ref{fig:UUG avgnum new old}. 
Two reasons account for this phenomenon roughly. 
One is that the old nodes are conducting less and less transactions in each window on average. The other is
the new nodes have few transactions (about 2) on average and their quantity is large. 
These two factors jointly lead to the decrease of average transactions as the time window moves forward.

\begin{figure}[h]
	\vspace{-0.4cm}
	\begin{minipage}[h]{0.24\textwidth}
		\centering
		\includegraphics[width=1\textwidth]{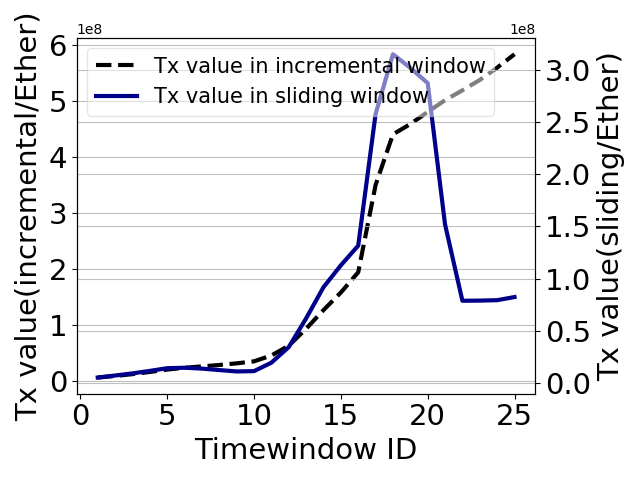}
		\caption{Total transaction value in UUG.}
		\label{fig:UUG sumvalue}
	\end{minipage}
	\begin{minipage}[h]{0.24\textwidth}
		\centering
		\includegraphics[width=1\textwidth]{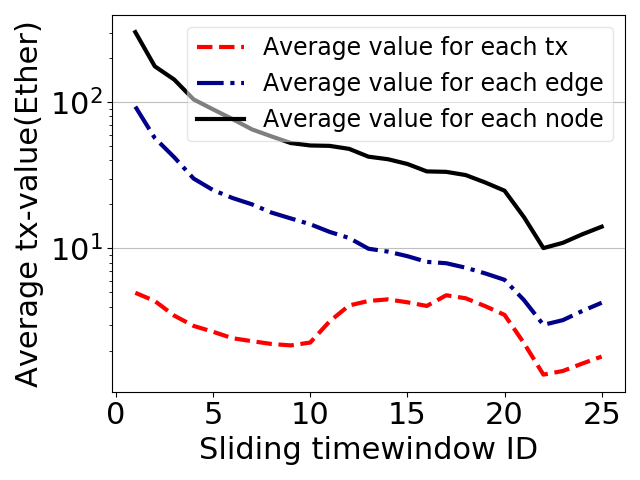}
		\caption{Average transaction value in UUG.}
		\label{fig:UUG average value}
	\end{minipage}
	\begin{minipage}[h]{0.24\textwidth}
		\centering
		\setlength\belowcaptionskip{-1pt}
		\includegraphics[width=1\textwidth]{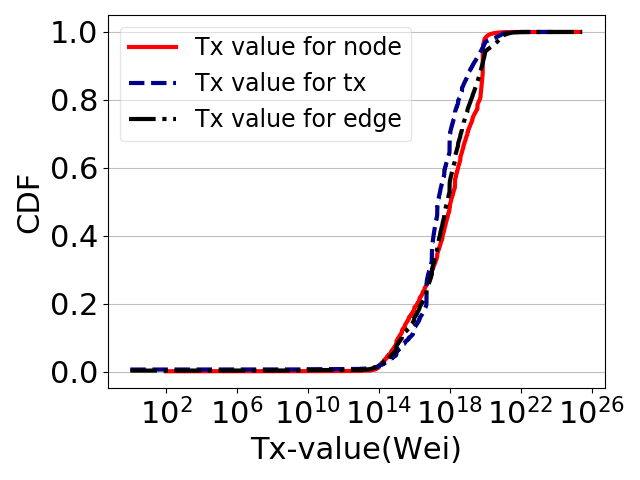}
		\caption{Transaction value distribution in UUG.}
		\label{fig:UUG value distribution}
	\end{minipage}
	\begin{minipage}[h]{0.24\textwidth}
		\centering
		\includegraphics[width=1\textwidth]{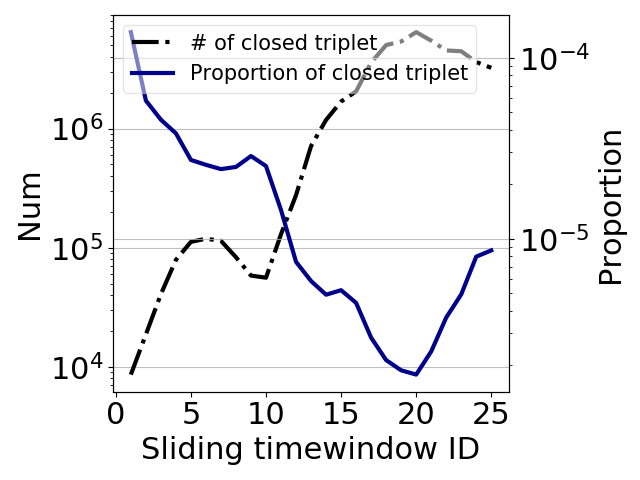}
		\caption{Number and proportion of closed triplet in UUG.}
		\label{fig:UUG motif circle sliding}
	\end{minipage}
	\vspace{-0.3cm}
\end{figure}

Ethereum, as a well-established cryptocurrency, has involved a huge amount of money. Hence, it is important to 
evaluate the value transfer on each transaction, each edge and each node. Ethereum adopts two currency units, \emph{Ether} and \emph{Wei}, which has ``1 Ether = $10^{18}$ Wei''. If not mentioned explicitly, we only use Ether. 
Figure \ref{fig:UUG sumvalue} plots the total value in circulation. 
In the sliding window case, the total value remains very steady before the tenth time window, and experiences a 
busty growth until the 18th window. An abrupt drop in the value is observed at the 20th window and it becomes 
more stable since the 22nd window. 
According to Figure \ref{fig:UUG average value}, the average value of each transaction 
decreases between the first window and the tenth window, and decreases from around 5 Ether to 2 Ether. 
The transaction values between the 11th window and the 19th window oscillate around 4 Ether. The transaction 
values drop below 1 Ether since the 21st window. Figure \ref{fig:UUG average value} further shows that the 
the average value transferred to or from each node and each edge is almost always decreasing over time (except the 22nd sliding window). We conjecture that two reasons lead to this phenomena. 
Firstly, the continually arriving new nodes usually have a few transactions, and the average value of each transaction decreases due to the increase of Ether price. 

The transaction value distribution of the up-to-date UUG in our dataset is shown in Figure \ref{fig:UUG value distribution}. It shows the value of most transactions, most nodes and most edges are all concentrated between $10^{14}$ Wei and $10^{21}$ Wei.

\noindent\textbf{Observation 3:} Around 83.63\% of edges have only a one-time transaction, and around 18.69\% of nodes trade only once.
The total number and value of transactions expriences three stages and the trend is consistent with price of Ether.
Average number and value of transactions decline over time, whether for each node or each edge and there is no evidence that the trend is price-related.

\subsection{Motifs of UUG}

The graph size and the degree distribution have portrayed the global property of Ethereum transaction networks. 
However, given the same global property, different networks may possess different local structures. These local 
structures usually are the basic building blocks of the entire graph that reveal the microscopic behaviors on how
the network is formed. In general, the subgraphs containing a few nodes and edges induced from the original graph, also 
named \emph{motifs} \cite{ref:motif}, are of particular interests to the network science research community.

\begin{figure}[h]
	\vspace{-0.4cm}
	\begin{minipage}[h]{0.5\textwidth}
		\centering
		\setlength\belowcaptionskip{-1pt}
		\includegraphics[width=1\textwidth]{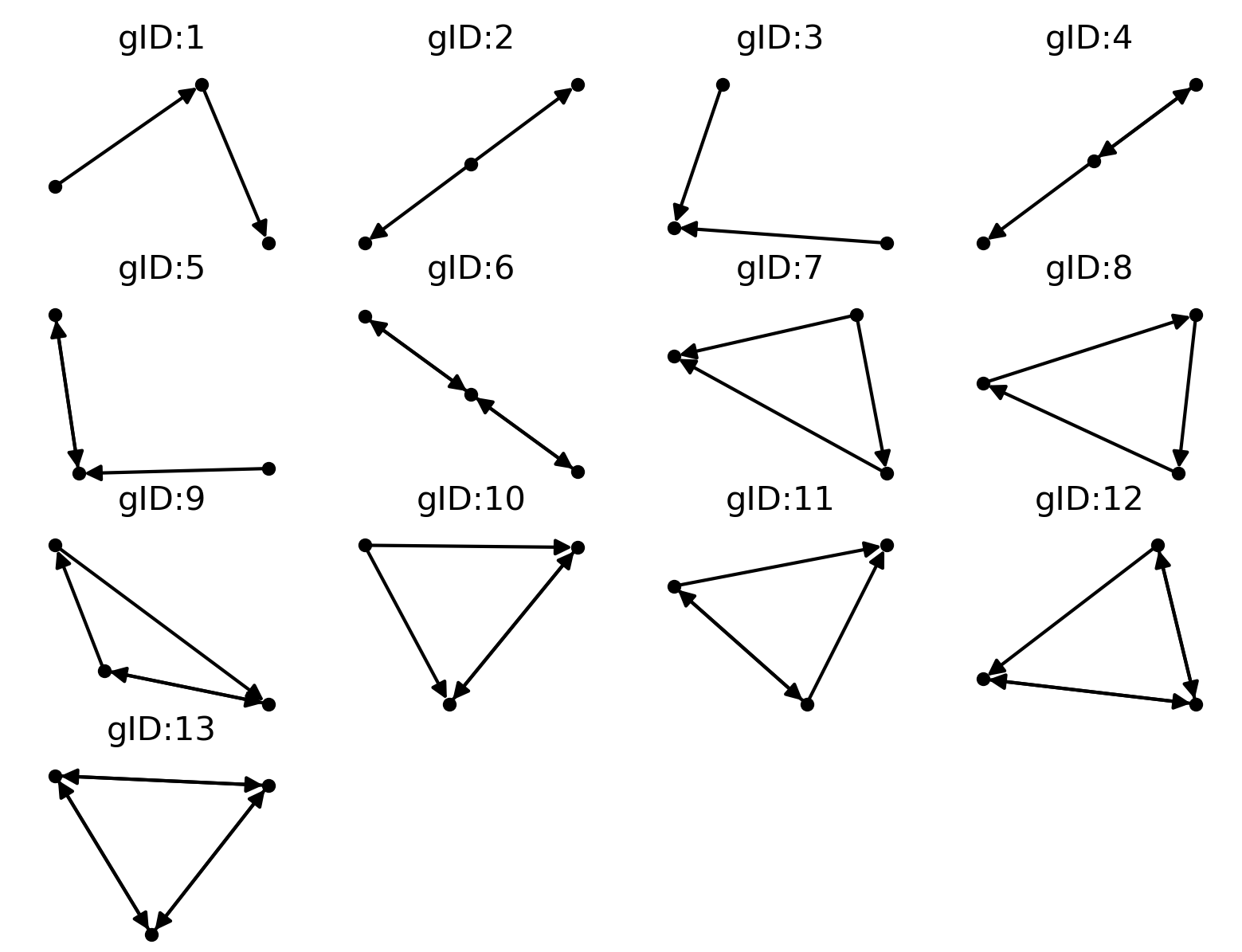}
		\vspace{-1cm}
		\caption{3-nodes motifs.}
		\label{fig:3 motif}
	\end{minipage}
	\vspace{-0.4cm}
\end{figure}
We hereby investigate the most important 3-node motifs of UUG. 
As a directed graph, there are thirteen motifs(Figure \ref{fig:3 motif}). We classify seven motifs as the 
\emph{closed triplet} that has transactions between each pair of nodes, and name the \emph{open triplet} 
for the remaining six motifs as a unity. The closed triplet symbols the closer transaction tiers in which 
more triplets represent more strong interactions among users. 
We illustrate the closed triplets in Figure \ref{fig:UUG motif circle sliding}. Here, $x$-coordinate is the sliding window, 
left $y$-coordinate is the number of the closed triplets and right $y$-coordinate is the proportion of the closed triplets over 
all the 3-node motifs. Note that we only consider the nodes that have transactions in each separately sliding window. 
In the very beginning, the number of closed triplets is around $10^{4}$, and it grows to around $4\times 10^{6}$ in the last 
sliding window. Although this number decreases considerably between the seventh window and the tenth window, the 
overall trend is the expansion of hundreds of times. This manifests that more and more EOAs are conducting 
transactions with the others. The proportion of closed triplets among all the 3-node motifs, unfortunately, is almost strictly 
decreasing from the beginning to the twentieth sliding window. The initial concentration is above ten thousandth 
(i.e.$10^{-4}$), it decreases to nearly the order of millionth (i.e.$10^{-6}$). The low concentration of the closed triplets 
means the transaction pattern of UUG is dominated by the mode which 
a vast majority of nodes merely interact with a small amount of nodes.

\begin{figure}[h]
	\vspace{-0.5cm}
	\begin{minipage}[h]{0.24\textwidth}
		\centering
		\includegraphics[width=1\textwidth]{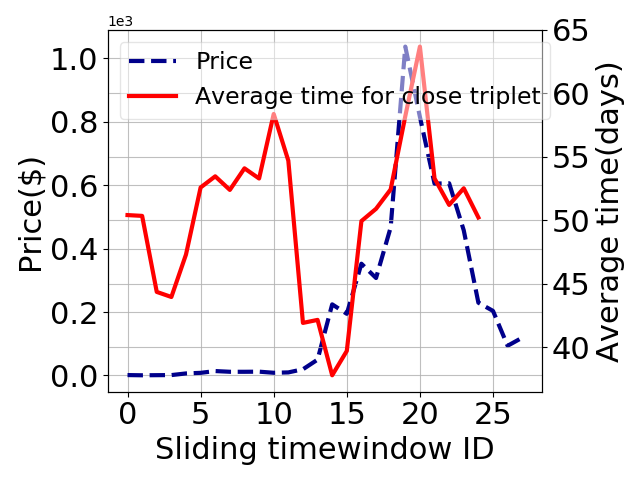}
		\caption{Average time for each closed triplet.}
		\label{fig:UUG avgtime motif}
	\end{minipage}
	\begin{minipage}[h]{0.24\textwidth}
		\centering
		\includegraphics[width=1\textwidth]{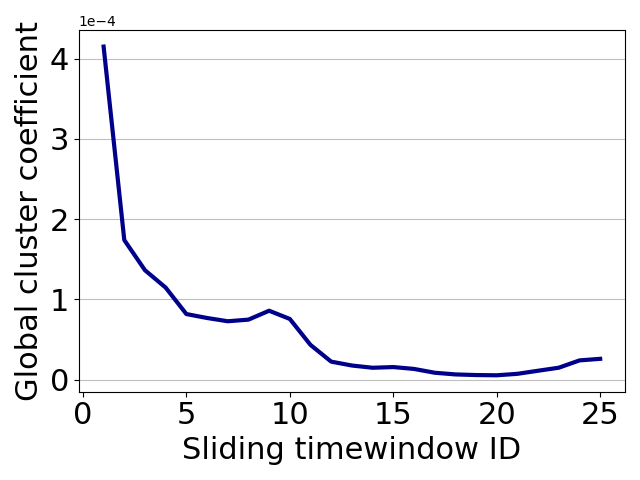}
		\caption{Global clustering coefficient evolution in UUG.}
		\label{fig:UUG global cluster sliding}
	\end{minipage}
	\vspace{-0.5cm}
\end{figure}

Another interesting question is how long time it will take to form a closed triplet. 
Figure \ref{fig:UUG avgtime motif} shows the average needed time for an open triplet to be closed. 
The average closure time varies at different sliding windows, i.e. ranging between 37 days and 
64 days, while their differences are not prominent, and most of the closure time oscillates between 50 days and 55 days. 
We further plot the dynamics of Ether price in Figure \ref{fig:UUG avgtime motif}. 
The correlation between the Ether price and the closure time is relatively weak. Especially, when the Ether price increases 
fifty to one hundred times, the closure time remains the same or experiences a relatively small increase. 

In graph theory, a clustering coefficient is a measure of the degree to which nodes in a graph tend to cluster together. 
In most of real-world networks, in particular online social networks, nodes tend to create 
tight-knit groups characterized by a relatively high density of ties; this likelihood tends to be greater than the average probability of a tie randomly established between two nodes.
Figure \ref{fig:UUG global cluster sliding} shows that the global clustering coefficient of UUG is usually small and even close to 0. The overall trend of the clustering coefficient is decreasing as the sliding window 
moves on. This manifest again that UUG is very sparse, and it is less likely for two trading nodes to form a strong closed triplet. 

\noindent\textbf{Observation 4: } The global cluster coefficient of Ethereum is very small. Although the number of closed triplet has increased, it is still negligible compared to the number of open triplet. The average closure time fluctuates at different time and there is no evidence that it affected by the price of Ether.

\section{Burstiness of Transactions in UUG}
\label{sec:burstiness}

The dynamic of Ethereum is driven by the loosely coordinated activity of a large 
number of users, economic and political factors.  
While we witnessed much progress in the study of Ethereum transactions, little is known 
about the dynamics characteristics. Burstiness is common temporal measure of 
the dynamics of various complex systems, and the burstiness of Ethereum's transaction patterns 
is the most important measure in connection with system dynamics as well. 
Here, we define and analyze the burstiness of Ethereum's transaction from both macroscopic and 
microscopic perspectives, in which the former gauges the active period of each user in his lifespan, 
and the latter looks into the more fine-grained time interval between consecutive transactions.

\subsection{Macroscopic Burstiness}

\begin{figure}[h]
	\vspace{-0.5cm}
	\begin{minipage}[!h]{0.24\textwidth}
		\centering
		\includegraphics[width=1\textwidth]{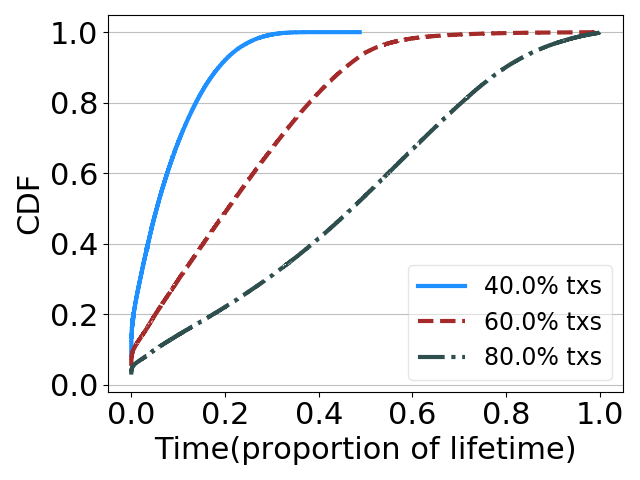}
		\caption{Busy period ratio distribution.}
		\label{fig:UUG interval distribution}
	\end{minipage}
	\begin{minipage}[!h]{0.24\textwidth}
		\includegraphics[width=0.8\textwidth]{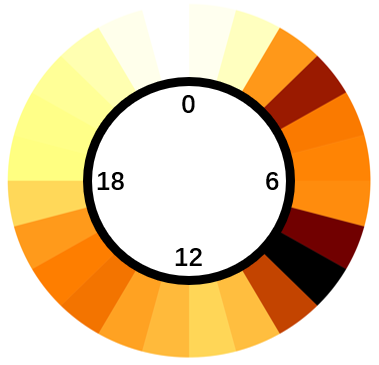}
		\caption{Burstiness of transactions in UUG during one day.}
		\label{fig:UUG burstiness global}
	\end{minipage}
	\vspace{-0.45cm}
\end{figure}

To quantify the macroscopic burstiness, we need to record the lifetime of each node. 
First, the lifetime of a node is defined as the time interval between the first transaction and the final transaction made by him. 
Our next step is to find the minimum time needed by each node to conduct a certain percentage of transactions. It is noting that the nodes with less than 10 transactions is eliminated. Too few transactions are meaningless for studying burstiness.
This time interval quantifies the most active or the bursty period for each node. 
Since the lifetimes of different nodes vary drastically, the absolute values cannot capture the statistics of burstiness of the 
whole population. Therefore, we define a new metric, the \emph{busy period ratio}, that is the least required time to 
complete a certain percentage of transactions divided by the lifetime of this node. 
Figure \ref{fig:UUG interval distribution} shows the results in three scenarios. 
One can observe that 80\% of nodes complete 40\% of the transactions with only 13.81\% of their lifetime; 80\% of nodes completed 60\% of transactions with 38.02\% of their lifetime, and 80\% of nodes use 70.46\% of their lifetime to complete 80\% of transactions. Our measurement shows the clear bursty evidence of almost all the meaningful nodes as a whole. 

The burstiness of global transactions is measured by the average number of transactions in each hour on a daily basis (GMT is adopted). Figure \ref{fig:UUG burstiness global} illustrates the intensity of bursty transactions made by 
EOAs. The most active period is  7\emph{am} $\sim$10\emph{am}, and the second one is 3\emph{am}$\sim$4\emph{am}. The most active period corresponds to 15 \emph{pm}$\sim$18\emph{pm} at Beijing Time (GMT+8), 
the afternoon working hours in Asian countries. Due to the escalating interests toward blockchains in these countries, we 
conjecture that this is the very reason accounting for the global burstiness in terms of the number of transactions in each day.

\subsection{Microscopic Burstiness}

The microscopic burstiness in Ethereum transactions refers to the significantly enhanced activity level over short periods of time followed by long periods of inactivity for a single node, similar to dozens of systems such as  email, web browsing, and
human contact patterns \cite{ref:bursty}. Goh and Barabasi proposed to characterize the (microscopic) bursty nature of 
social events using orthogonal measures \cite{ref:MB}. They quantify two distinct mechanisms causing burstiness: the inter-event time 
distribution and the \emph{memory}. With a positive memory, the short inter-event times are prone to following short ones, while 
with a negative memory, the short (resp. long) inter-event times tend to be followed by long (short) ones. Inspired by this 
pioneering study, we adopt the same method to examine the burstiness and memory of Ethereum transactions.

\begin{figure}[!h]
	\vspace{-0.5cm}
	\begin{minipage}[!h]{0.24\textwidth}
		\centering
		\setlength\abovecaptionskip{-0.5pt}
		\setlength\belowcaptionskip{-1pt}
		\includegraphics[width=1\textwidth,height=0.15\textheight]{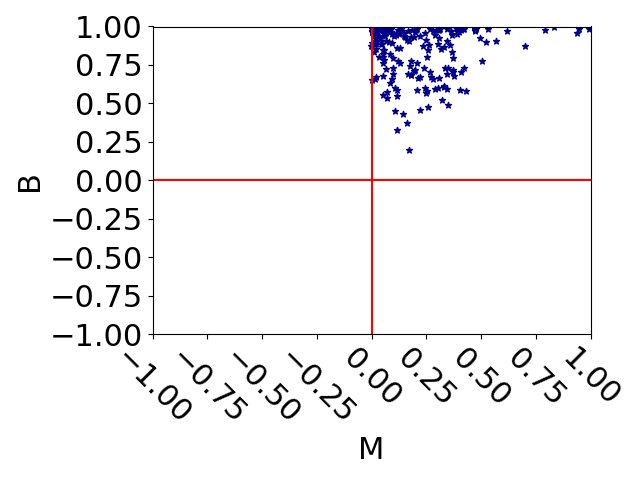}
		\caption{MB of normal nodes.}
		\label{fig:UUG MB normal}
	\end{minipage}
	\begin{minipage}[h]{0.24\textwidth}
		\centering
		\setlength\abovecaptionskip{-0.5pt}
		\setlength\belowcaptionskip{-1pt}
		\includegraphics[width=1\textwidth,height=0.15\textheight]{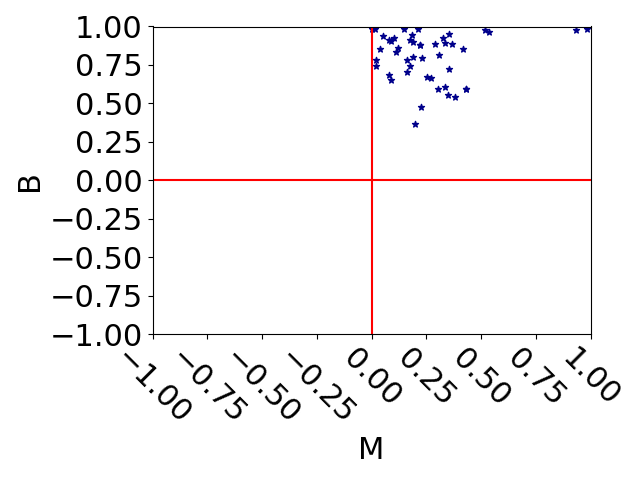}
		\caption{MB of exchanges.}
		\label{fig:UUG MB exchange}
	\end{minipage}
	\begin{minipage}[!h]{0.24\textwidth}
		\centering
		\setlength\abovecaptionskip{-0.5pt}
		\setlength\belowcaptionskip{-1pt}
		\includegraphics[width=1\textwidth,height=0.15\textheight]{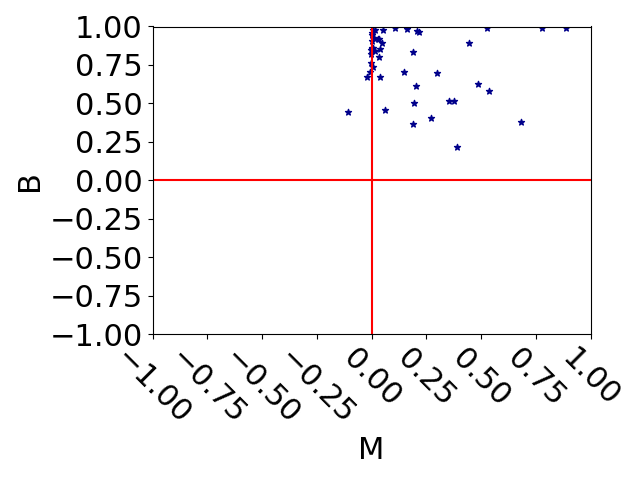}
		\caption{MB of mining pools.}
		\label{fig:UUG MB mining}
	\end{minipage}
	\begin{minipage}[!h]{0.24\textwidth}
		\centering
		\setlength\abovecaptionskip{-0.5pt}
		\setlength\belowcaptionskip{-1pt}
		\includegraphics[width=1\textwidth,height=0.15\textheight]{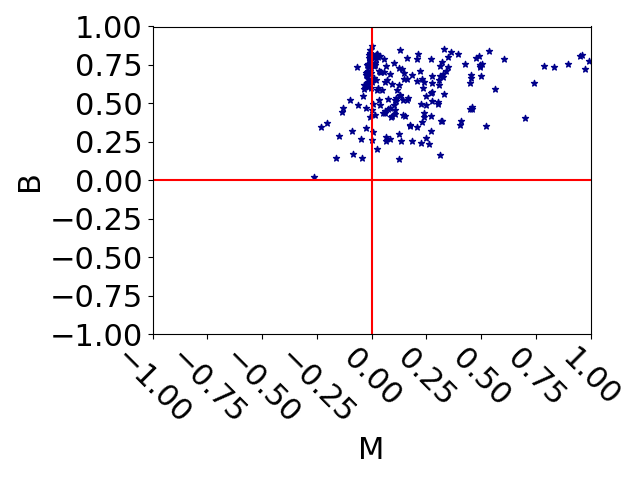}
		\caption{MB of phish hack.}
		\label{fig:UUG MB phish}
	\end{minipage}
	\vspace{-0.5cm}
\end{figure}

Similarily we define \emph{B} as the burstiness coefficient. \emph{B} is used to measure the inter-event time distribution  and it is defined as the function of the coefficient of variation of inter-event times.
\begin{equation}
\setlength\abovedisplayskip{1pt}
\setlength\belowdisplayskip{1pt}
\mathbf {\emph{B}} \equiv \frac{\sigma-<\tau>}{\sigma+<\tau>}
\label{eq:B}
\end{equation}
$<\tau>$ and $\sigma$ represent the mean and standard deviation of inter-event times.
Here \emph{B} takes the value of -1 for regular time series with $\sigma$=0, and it is equal to 0 for random, Poissonian time series where $\sigma=<\tau>$. 

Goh and Barab\'asi \cite {ref:MB} introduced the memory coefficient \emph{M} to measure two-point correlations between consecutive inter-event times as the following:
\begin{equation}
\setlength\abovedisplayskip{1pt}
\setlength\belowdisplayskip{1pt}
\mathbf {\emph{M}} \equiv  \frac{1}{n-2}
\sum_{i=1}^{n-2}
\frac{((\tau_i)-\left<\tau\right>_1)(\tau_{i+m}-\left<\tau\right>_2)}{(\sigma_1)(\sigma_2)}
\label{eq:M}
\end{equation}
with $\left<\tau\right>_1$ (respectively$\ \left<\tau\right>_2$) and $\sigma_1$ (respectively $\sigma_2$) being the average and the standard deviation of inter-event times {$\{\tau_i|i=1,\cdots,n-2\}$}(respectively {$\{\tau_{i+1}|i=1,\cdots,n-2\}$}). The closer M is to 1, the greater the probability that the time interval after the long (short) interval is long (short). The closer M is to -1, the greater the probability that the time interval after the long (resp. short) interval is short (resp. long).
\vspace{0.1cm}

The same number of accounts are selected from normal accounts, exchanges, mining pools and fraud accounts respectively, and then we map the transaction of these accounts on the (\emph{M},\emph{B}-space). Figure \ref{fig:UUG MB exchange} and Figure \ref{fig:UUG MB mining} show the transactions of exchanges and mining pools are all busty, we conjecture this may be caused by the burstiness of transactions during on day. Compared with Figure \ref{fig:UUG MB normal}, Figure \ref{fig:UUG MB phish} shows the memory of phishing accounts are distributed on a wider range.

\noindent\textbf{Observation 5: }From perspective of macroscopic burstiness, the transactions of most of nodes are concentrated on a small part of the lifetime of the nodes and the transactions during one day is distributed on some hours. From perpective of microscopic burstiness, the arrival time distribution of transactions of nodes is very highly divise.  
	\section{The Rich Gets Richer in UUG}
The Matthew effect can be observed in many aspects of social and economic systems. 
It is sometimes known as the adage ``the rich get richer and the poor get poorer''. 
A natural question arises with regard to (w.r.t.) Ethereum: ``Is the rich getting richer?'' \cite{ref:rich get richer}.
It is crucial to know whether Ethereum will evolve into an extremely unhealthy economic system. 

In Ethereum transaction networks, we measure the Matthew effect from three aspects, namely degree, transaction number, and balance. The metric of assessing the Matthew effect is the Gini coefficient, a commonly used measure 
for the income gap. The Gini coefficient is in the range $[0, 1]$: $[0,0.2)$ represents absolute average income, $[0.2,0.4)$ represents the income is relatively reasonable; $[0.4,0.5)$ represents the income gap is large, $[0.5,1]$ means that income is too unbalanced.
\subsection{Degree Geni}

\begin{figure}[!h]
	\vspace{-0.3cm}
	\begin{minipage}[!t]{0.24\textwidth}
		\centering
		\includegraphics[width=1\textwidth]{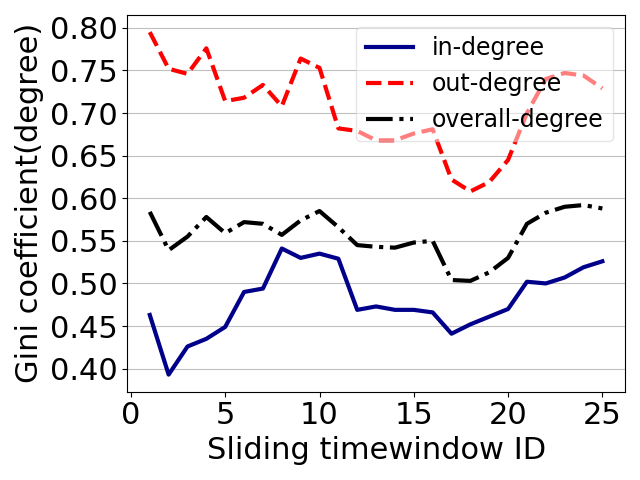}
		\caption{Gini of UUG degree in sliding windows.}
		\label{fig:UUG gini degree sliding}
	\end{minipage}
	\begin{minipage}[!t]{0.24\textwidth}
		\centering
		\includegraphics[width=1\textwidth]{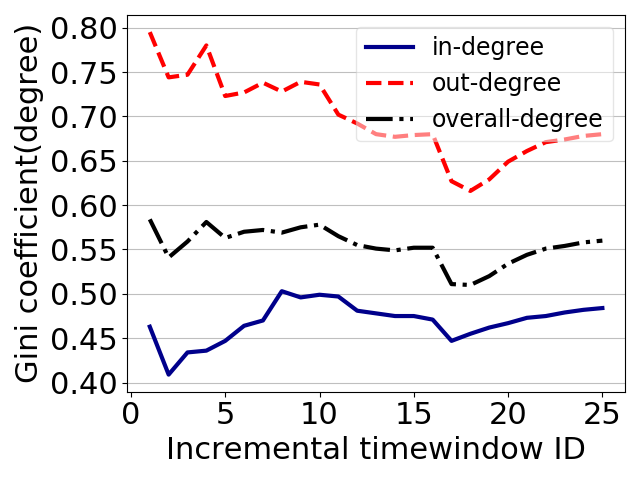}
		\caption{Geni of UUG degree in incremental windows.}
		\label{fig:UUG gini degree incremental}
	\end{minipage}
	\vspace{-0.3cm}
\end{figure}

The Gini coefficients of the UUG degree are shown in Figure \ref{fig:UUG gini degree sliding} with sliding windows. 
We discover that the Gini coefficient is above 0.4 whichever the degree, the in-degree or the out-degree is 
used. It means the degree distribution is unbalance no matter in which time period. The Gini coefficient of the in-degree is smaller than that of the out-degree both in sliding and incremental graphs, meaning that the distribution of out-degree is more unbalance. 
The Gini coefficients of the overall degree fluctuates in the range $[0.5, 0.6]$ when the 
incremental window is considered. But there is no sign on the direction that the Gini coefficient is moving toward. 
Hence, we can conclude that in terms of degree distribution, ``Always unfair but not the rich gets richer.''

\begin{figure}[t]
	\begin{minipage}[t]{0.24\textwidth}
		\includegraphics[width=1\textwidth]{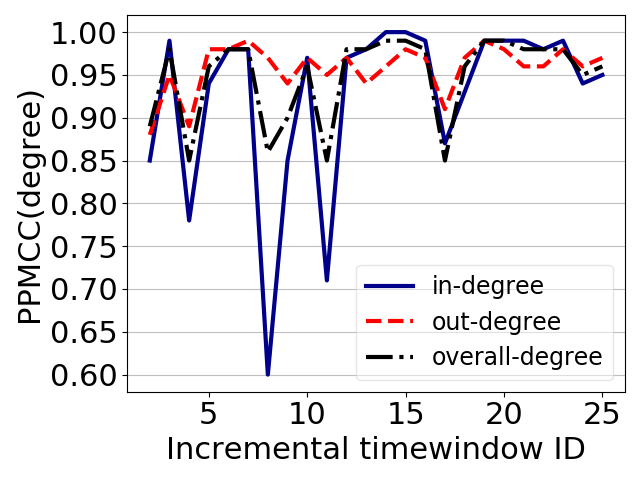}
		\caption{PPMCC of UUG degree in incremental windows.}
		\label{fig:UUG pearson degree incremental}
	\end{minipage}
	\begin{minipage}[t]{0.24\textwidth}
		\centering
		\includegraphics[width=1\textwidth]{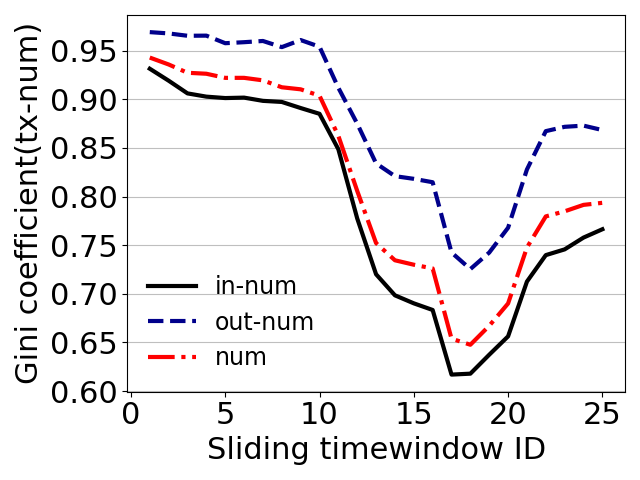}
		\caption{Gini coefficient of UUG transaction number in sliding windows.}
		\label{fig:UUG gini num sliding}
	\end{minipage}
	\vspace{-0.3cm}
\end{figure}

A subsequent question is ``will the rich still be rich in the near future?''.
In Gini coefficient, there is no differentiation on the node identity. 
We then introduce the Pearson product-moment correlation coefficient (PPMCC) to explore this question. 
In statistics, PPMCC is a measure of linear correlation between two variables, and it takes a value between -1 and 1. 
A positive (resp. negative) PPMCC indicates the positive (resp. negative) linear correlation. When PPMCC is 0, 
it means that there is no linear correlation. More specifically, there is a very strong correlation 
with $PPMCC\in(0.8, 1]$, a strong one with $PPMCC\in(0.6, 0.8]$, a moderate one 
with $PPMCC\in(0.4, 0.6]$, a weak one with $PPMCC\in(0.2, 0.4]$, a weak or no correlation with $PPMCC\in[0, 0.2]$.
We evaluate PPMCC of the degree of 
nodes in consecutive windows in Figure \ref{fig:UUG pearson degree incremental}. 
This implies that PPMCC is very large between consecutive windows, which means that the nodes with rich degrees 
are still rich, and those with poor degrees are still poor.


\begin{figure}[!h]
	\begin{minipage}[!h]{0.24\textwidth}
		\centering
		\setlength\abovecaptionskip{-0.5pt}
		\setlength\belowcaptionskip{-1pt}
		\includegraphics[width=1\textwidth]{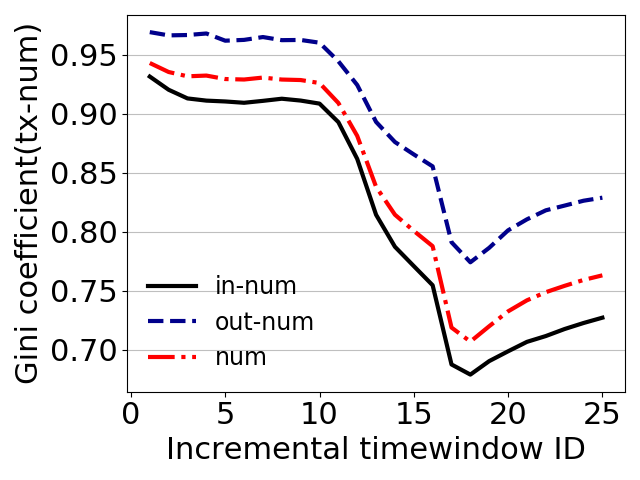}
		\caption{Gini coefficient of UUG transaction number in incremental windows.}
		\label{fig:UUG gini num incremental}
	\end{minipage}
	\begin{minipage}[!h]{0.24\textwidth}
		\setlength\abovecaptionskip{-0.5pt}
		\setlength\belowcaptionskip{-1pt}
		\includegraphics[width=1\textwidth]{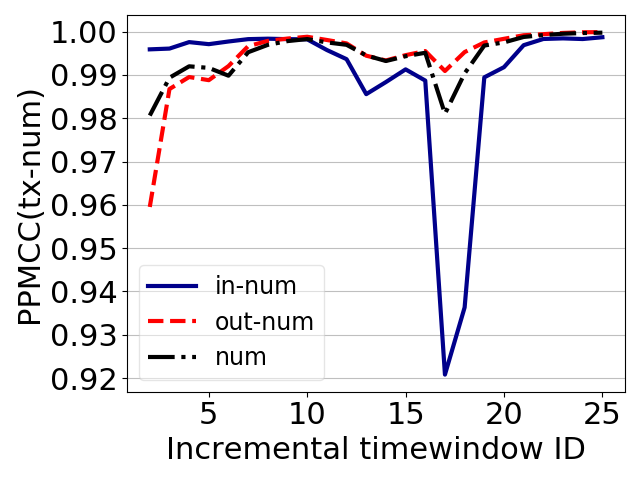}
		\caption{PPMCC of UUG transaction number  incremental windows.}
		\label{fig:UUG pearson num incremental}
	\end{minipage}
	\vspace{-0.5cm}
\end{figure}

\subsection{Transaction Gini}

We next examine whether the rich gets richer in terms of the number of transactions. 
Similarly, the Gini coefficient is adopted to quantify the concentration of transactions among different nodes. 
Figure \ref{fig:UUG gini num sliding} shows that the Gini coefficient is above 0.6 in every sliding time window. 
Such an observation exists regardless of receiving (e.g. in-num) and transferring (e.g. out-num) Ethers. 
According to the well-recognized criterion, a Gini coefficient more than 0.6 is deemed as an very unfair system. 
The Gini coefficient experiences a sharp drop from the tenth to the seventeenth sliding windows, and increases fast 
afterwards. We conjecture that the decrease of Gini coefficient is due to the ascending popularity of Ethereum, and 
the increase of Gini coefficient is related to the rise of Ether price. 
We also observe that the unfairness in transferring Ether is even more severe than in receiving Ether. 
Figure \ref{fig:UUG gini num incremental} illustrates the Gini coefficients in the incremental time window. 
The extent of unfairness in the number of transactions, though lower than that in the sliding time window, is still very serious.  
Therefore, we observe that in terms of transaction number distribution, ``Always very unfair but not the rich get richer''. 

The question``will the rich still be rich in the near future?" is also studied in terms the number of transactions. Figure \ref{fig:UUG pearson num incremental} shows PPMCC is greater than 0.98 in most time periods. This shows there is a very strong correlation in number of transactions in different time periods. Similar with degree, we can conclude that nodes with rich transaction number are still rich at the next moment.

\subsection{Balance Gini} 

As a crypto currency system, Ethereum sees hundreds of thousands of currency circulations every day on average. 
Meanwhile, new Ethers are created nearly every 15 seconds. It is unclear how the Ethers 
are accumulated and shared among all the nodes. We hope that Geni coefficient of nodes' balances 
can reveal the (un)fairness of wealth distribution in Ethereum. 

\begin{figure}[h]
	\vspace{-0.5cm}
	\begin{minipage}[h]{0.24\textwidth}
		\centering
		\includegraphics[width=1\textwidth]{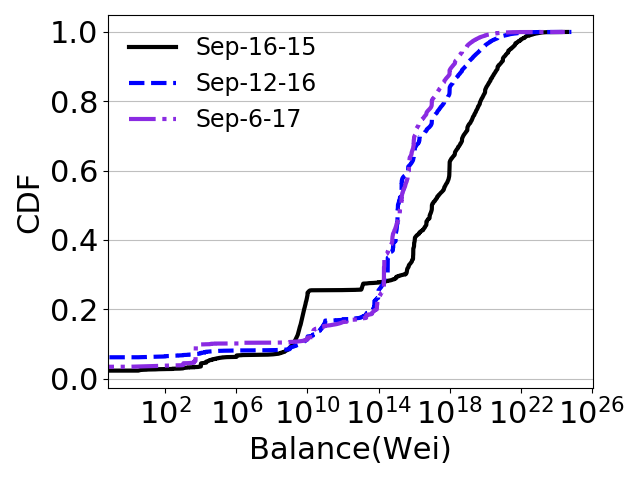}
		\caption{Balance distribution in UUG.}
		\label{fig:UUG balance distribution}
	\end{minipage}
	\begin{minipage}[h]{0.24\textwidth}
		\centering
		\includegraphics[width=1\textwidth]{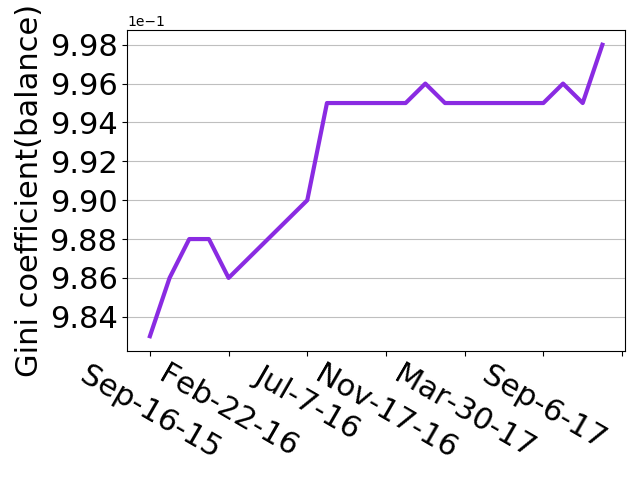}
		\caption{Gini coefficient of UUG balance.}
		\label{fig:UUG gini balance mvdie}
	\end{minipage}
	\vspace{-0.3cm}
\end{figure}

As the first step, we assume that a node is ``dead'' if his balance is 0, and he does not trade with any others afterwards. 
Figure \ref{fig:UUG balance distribution} plots the balance distribution of all living users at different sampling time. 
One can see that around 62\% of nodes possess a balance below $10^{18}$ Wei on September 16, 2015, 
while around 90\% of nodes are below this level on September 12, 2016 and September 06, 2017. 
This implies that the Ether wealth is diluted to more nodes.

We then calculate the Gini coefficient of nodes' balances in Figure \ref{fig:UUG gini balance mvdie}. At the initial phase of 
Ethereum, the Gini coefficient of balance is 0.982, an astonishingly high value. As time moves on, it almost increases to 
0.998. This is to say, Ethereum is ``extremely unfair in terms of the balance since its birth''. 

\begin{figure}[!h]
	\begin{minipage}[!h]{0.24\textwidth}
		\vspace{-0.4cm}
		\centering
		\includegraphics[width=1\textwidth]{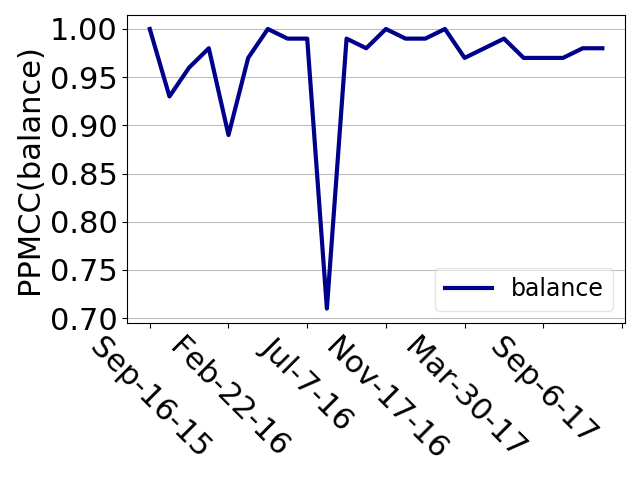}
		\caption{PPMCC of UUG balance.}
		\label{fig:UUG pearson balance}
	\end{minipage}
	\begin{minipage}[!h]{0.24\textwidth}
		\vspace{-0.4cm}
		\centering
		\includegraphics[width=1\textwidth]{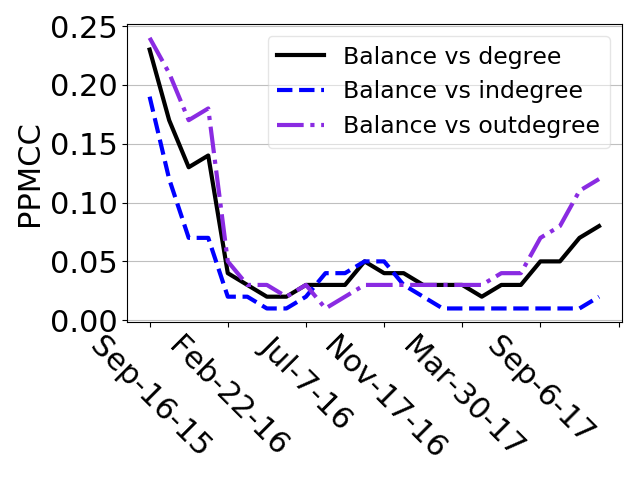}
		\caption{PPMCC of UUG degree and balance.}
		\label{fig:UUG pearson degree balance}
	\end{minipage}
	\vspace{-0.5cm}
\end{figure}

We further explore the factors that affect the balance of users. Pearson coefficient is utilized 
to measure the relationship between the balance at different time. Figure \ref{fig:UUG pearson balance} shows 
that the Pearson coefficient is very large, i.e. above 0.85 except for a few moments. 
This means that the rich remains to be rich at next moment. 200 thousand blocks is selected as the measurement interval (about 34 days).
Similarly, the Pearson coefficient is used to judge the relationship between the balance and the degree. 
As shown in Figure \ref{fig:UUG pearson degree balance}, the results range between 
0.01 and 0.25, and at most time is below 0.1. 
The results show that for the same user, there is not much correlation between the balance and degree.

\noindent\textbf{Observation 6: }The distribution of degree and transaction number are always unfair but not the rich get richer. The distribution of wealth is extremely unfair since the birth of Ethereum.

\section{Temporal UCG and CCG Network Analysis}
Compared with Bitcoin, the biggest difference in Ethereum is that it can support a more powerful scripting language, allowing developers to develop arbitrary applications on it and implement arbitrary smart contracts. We structure the transactions involving smart contracts into two diagrams, UCG and CCG, for comparison and introduction next. 

Three operations of contracts are explored, including ``create()", ``call()" and ``suicide()". The differences of each operation in different periods are studied.
\subsection{Creation of contracts.}
\begin{figure}[h]
	\vspace{-0.5cm}
	\begin{minipage}[h]{0.24\textwidth}
		
		\centering
		\includegraphics[width=1\textwidth]{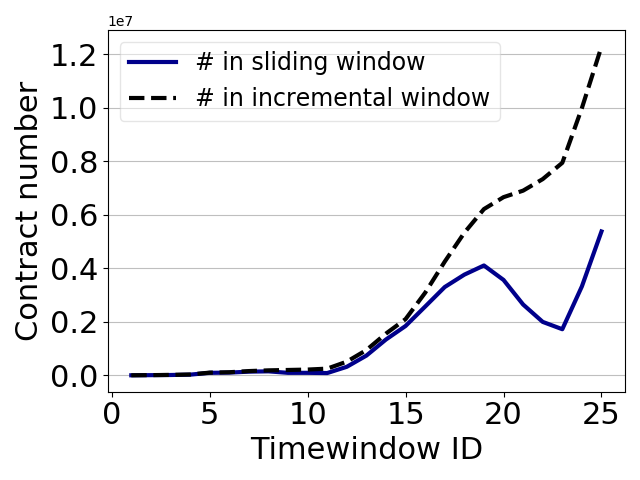}
		\caption{Number of contracts are created.}
		\label{fig:contract sumnum}
	\end{minipage}
	\begin{minipage}[h]{0.24\textwidth}	
		\centering
		\setlength\abovecaptionskip{-0.5pt}
		\setlength\belowcaptionskip{-1pt}
		\includegraphics[width=1\textwidth]{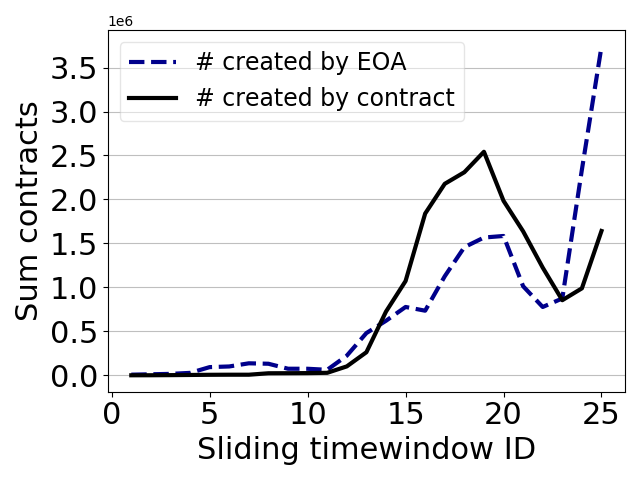}
		\caption{Number of contracts in each timewindow.}
		\label{fig:contract creat divide sliding}
	\end{minipage}
	\vspace{-0.6cm}
\end{figure}

Contracts can be created by EOAs or contracts.  
The development of smart contracts also experiences three stages from Figure \ref{fig:contract sumnum}. It is different that the length of ``outbreak" is shorter and in last time window the number becomes much higher.  
Figure \ref{fig:contract creat divide sliding} shows the re-boom is mainly caused by EOAs. There is little difference between the number of contracts which is created by EOAs and contracts except the last period.

\subsection{Call of contracts.}
The code of smart contracts runs distributed in the Ethereum virtual machine at each node in the network. Call smart contracts is to initiate a transaction that points to the smart contract address. The initiator can be EOA or other contracts.

\makeatletter\def\@captype{figure}\makeatother
\begin{minipage}[h]{0.24\textwidth}	
	\centering
	\setlength\abovecaptionskip{-2pt}
	\setlength\belowcaptionskip{-1pt}
	\includegraphics[width=1\textwidth]{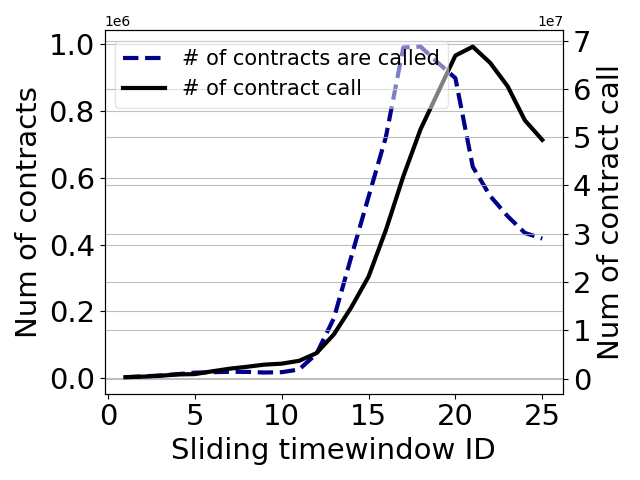}
	\caption{Number of contracts are called and contract call.}
	\label{fig:sumcall sliding}
\end{minipage}
\makeatletter\def\@captype{table}\makeatother
\captionsetup{singlelinecheck=off} 
\begin{minipage}[h]{0.22\textwidth}
	\begin{tabular}{|p{0.8cm}|p{1.3cm}|p{0.8cm}|}
		\hline  
		{Token}&{Exchange}&{Others}\\
		\hline 
		{$10$}&{$7$}&{$3$}\\
		\hline  
	\end{tabular}
	\caption {Top 20 contrasts with most call.}
	\label{table:top contrast}
\end{minipage}
\vspace{-0.1cm}
\begin{figure}[h]
	\vspace{-0.3cm}
	\begin{minipage}[t]{0.24\textwidth}
		
		\centering
		\includegraphics[width=1\textwidth]{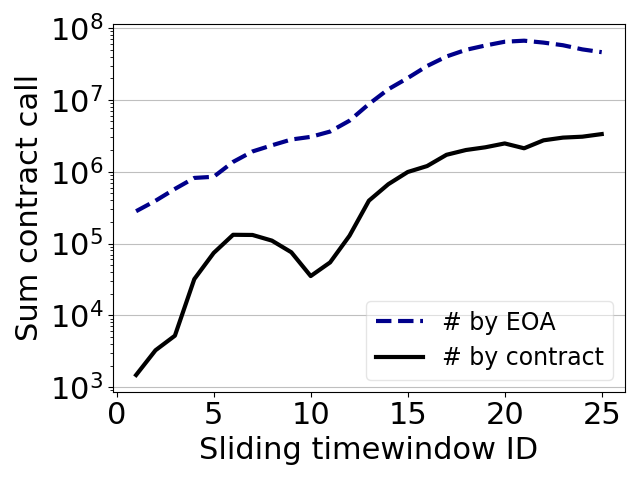}
		\caption{Number of contract call by EOAs and contracts.}
		\label{fig:sumcall divided sliding}
	\end{minipage}
	\begin{minipage}[t]{0.24\textwidth}
		
		\centering
		\includegraphics[width=1\textwidth]{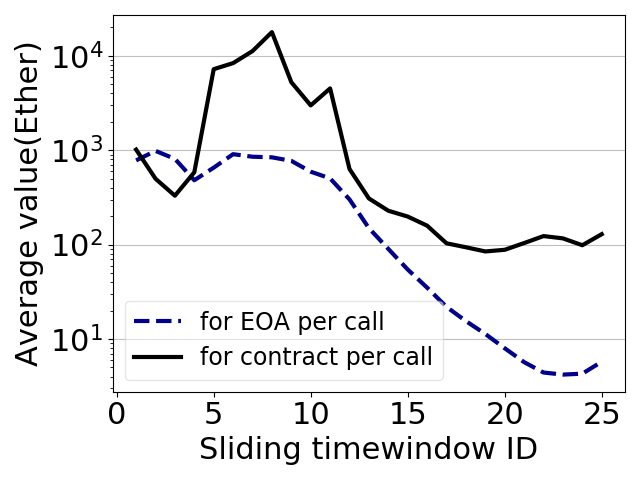}
		\caption{Average value for each call.}
		\label{fig:average value call}
	\end{minipage}
	\vspace{-0.3cm}
\end{figure}
Figure \ref{fig:sumcall sliding} shows that the number of contracts which are called and contract invocations all experience three-stage similar with the development of other elements. Figure \ref{fig:sumcall divided sliding} shows EOAs call contracts much more times than contracts. We can conclude that the contracts are mainly invoked by EOAs. Table \ref{table:top contrast} shows contracts with most call are mainly token and exchange contract. Figure \ref{fig:average value call} shows that a smart contract spends more Ethers than an EOA on average when they call a contract.

\subsection{Suicide of contracts.}
Smart contract supports contract suicide function that is programed to tenminate the contract upon certain conditions. When the contract suicides, the remaining Ether associated with this contract  will be transferred to another address. We find that in our dataset, there are 7,995,732 destruction contracts that transfer money to EOA and only two suicide contracts transfer money to other contracts.

\noindent\textbf{Observation 7: } The development of smart contract can be divided into three stages like the evolution of degree and transaction numbers. The number of contracts created by EOAs and contracts is of the same order of magnitude, but the contracts are mainly called by EOAs. When the contracts suicide, most of them transfer the balance to EOAs.  
	\section{Conclusion}
We conduct an evolution of Ethereum from the perspective of temporal graph analysis. The data analytics platform is developed to collect external transactions and internal transactions. These transactions are constructed into three graphs based on the trading relationship and they are compared among different time windows. By analyzing these graphs through various metrics, we obtain many new observations and insights, which help people have a understanding of the evolution of Ethereum. We also explored the role of Ether price in the development of Ethereum. Moreover, we conduct research on the distribution of arrival time of transactions. The macroscopic and microscopic burstiness is validated. In addition, the distribution of wealth and other transaction indicators is always unfair throughout the development.

\end{document}